# Phonon limited mobility of the Dirac-Fermions in a three-dimensional Dirac semimetal Cd$_3$As$_2$

S S Kubakaddi[*]

Department of Physics, K L E Technological University, Hubballi -580031, Karnataka, India.

**Abstract**

A theoretical model is presented for the phonon limited mobility of the Dirac-Fermion gas in a three-dimensional (3D) Dirac semimetal Cd$_3$As$_2$ considering scattering from both the acoustic and the optical phonons. Screening effects are taken into account and it is found that they lead to a significant enhancement of the mobility. Simple analytical equations and power laws are obtained in both the Bloch-Gruneisen and equipartition regimes. The dependence of mobility on the temperature $T$ and the electron density $n_e$ is investigated. It is found that optical phonon limited mobility $\mu_{op}$ dominates over the acoustic phonon limited mobility $\mu_{ap}$ in the higher temperature region. There is a crossover of $\mu_{ap}$ and $\mu_{op}$ and the crossover temperature $T_c$ shifts to higher value with increasing $n_e$. Our calculations of the mobility are in good agreement with the recent experimental data. Comparison is also made with the results in the conventional 3D electron gas in a degenerate semiconductor.

*email: sskubakaddi@gmail.com

Key words: Three-dimensional Dirac semimetal, mobility, electron-phonon coupling.

PACS No.s: 71.55.Ak, 72.10.-d, 72.10.Di, 72.20.Fr

**I. INTRODUCTION**

In recent years, there has been a great surge in the study of three-dimensional (3D) Dirac semimetal (3DDS) cadmium arsenide Cd$_3$As$_2$ [1-22], the 3D analogue of graphene. Due to the linear band structure of 3D Dirac-Fermions 3DDS Cd$_3$As$_2$ exhibits many unusual transport phenomena such as strong quantum oscillations [7, 13], ultrahigh mobility [10,11,13] and giant magnetoresistance [9,10,11, 14]. In order to find the potential applications of 3DDS Cd$_3$As$_2$ in high speed devices, it is essential to have a good understanding of its electronic transport properties, in particular of its mobility. One of the most fascinating features of 3DDS Cd$_3$As$_2$ is the observed ultra-high mobility, which is attributed to the suppressed back scattering of high velocity 3D Dirac-Fermions. Secondly, the carrier density $n_e$ is ultra-large and it is in the range of ~$10^{17}$ - $10^{20}$ cm$^{-3}$ [5,12,23-25]. The reported low temperature (~5 K) high mobility ~ $9 \times 10^6$ cm$^2$/V-s [5, 6,10,13] and up to $4.60 \times 10^7$ cm$^2$/V-s [13], resulting from the suppression of backscattering, are higher than the mobility in suspended graphene and the highest mobility of any bulk semiconductor.

From the Hall and resistivity data in Ref. [9], the transport mobility is estimated to be $\mu = 8 \times 10^4$ cm$^2$/V-s for the sample with $n_e =1.8 \times 10^{18}$ cm$^{−3}$. At higher temperatures the onset of phonon scattering is predicted [9]. Zhao et al. [13] have observed decreasing resistivity $\rho$, almost linearly, when $T$ decreases from 300 K to about 6 K. Then, for $T < 6$ K it tends to saturate at a quite low value of about 11.6 n$\Omega$-cm. The observed linear behavior is inferred to be due to the umklapp processes and the electron scattering by optical phonons. From the measured $\rho$ of about 11.6 n$\Omega$-cm at 6 K, the mobility $\mu$ (= $1/ n_e e \rho$) is predicted to be $9.18 \times 10^7$ cm$^2$/V-s for $n_e$=$5.86 \times 10^{18}$ cm$^{-3}$ [13]. In Refs. [10, 13], the values of residual resistivity of the samples appear to be negligibly small down to ~5 K.

In a 3DDS Cd$_3$As$_2$ sample of Cao et al. [12], the temperature dependence of the longitudinal resistivity $\rho_{xx}$, at zero magnetic field, gives $\rho_{xx}$ ~$T$ (i.e. $\mu \sim T^{-1}$), which is a typical metallic behavior, due to its semimetal band structure. They have also studied the temperature dependence of mobility. It is found that $\mu$ increases to the value of $1.9 \times 10^5$ cm$^2$/ V-s at 2.6 K, due to the reduced phonon scattering at very low temperatures. Further, the residual resistivity and the corresponding Hall mobility are 20 $\mu\Omega$- cm and $1.9 \times 10^5$ cm$^2$/ V-s, respectively. The carrier density of their sample exhibits a small change with temperature and reaches a relatively low value of $n_e$=$1.57 \times 10^{18}$ cm$^{-3}$ at 300 K. Most of their other Cd$_3$As$_2$ samples have high electron mobility in the range of 1-5 $\times 10^4$ cm$^2$/ V-s at room temperature. Pariari et al. [23] have also observed, for a Cd$_3$As$_2$ sample with $n_e \sim 6.8 \times 10^{18}$ cm$^{-3}$, a weak metallic behavior $d\rho/dT > 0$ over the entire temperature range from 350 to 2 K with an estimated mobility $\mu \sim 1.3 \times 10^4$ cm$^2$/V-s ($\rho \sim 70.612 \mu\Omega$-cm) at 2 K. Recently, metallic behavior has been also observed over a wide temperature range of 50-500 K with residual resistivity of $\sim 0.5$ $\mu\Omega$-m below 50 K [21].

High quality 3DDS Cd$_3$As$_2$ microbelts [26] and nanobelts [27], with the room temperature electron mobility $\sim 2 \times 10^4$ cm$^2$/V-s, have been fabricated paving the way for exciting electronic applications. In nanobelt, the Hall mobility $\mu_H$ follows the typical relation $\mu_H \sim T^{-\gamma}$ with $\gamma = 0.5$ for 20 K $< T <$ 200 K [27]. This decrease of $\mu_H$ with rise in $T$ is ascribed to the enhanced electron-phonon scattering.

There are other experimental evidences showing the importance of electron-phonon interaction. For example, the hot electron cooling of photoexcited carriers, mediated by phonons, has been investigated in 3DDS Cd$_3$As$_2$ by pump-probe measurements [17, 28, 29]. It is found to be dominated by electron coupling with acoustic phonons and relatively low energy optical phonons.

From the above experimental observations of transport properties, we notice that the metallic behavior is common in all the 3DDS Cd$_3$As$_2$ samples at relatively high temperature and sometimes down to $\sim$5 K. The source of such behavior is mainly the electron scattering by phonons. Since the mobility $\mu$ of the carriers governs the speed of the electronic devices, the understanding of the electron-phonon interaction in Cd$_3$As$_2$ is of fundamental importance for designing devices such as high mobility transistors. The electron-phonon (el-ph) interactions limit the intrinsic mobility and other transport properties at relatively higher temperatures.

There exist theoretical studies of some of the transport properties involving electron-acoustic phonon (el-ap) interaction in 3DDS Cd$_3$As$_2$ such as momentum relaxation time [30], hot electron cooling [31, 32], phonon-drag thermopower [33] and Cerenkov emission [34]. In addition, the steady state hot electron cooling power $P$ due to the polar optical phonon (el-op) scattering has been investigated [35]. In the study of momentum relaxation time [30], scattering due to acoustic phonon coupling is briefly addressed. The $P$ calculations show the dominance of the scattering by optical phonons for $T > \sim 25$ K [35]. However, there are no studies so far of the electron momentum relaxation and mobility due to scattering by optical phonons in 3DDS Cd$_3$As$_2$, which is expected to govern the transport at higher temperatures.

In the present work we theoretically study the phonon limited electron mobility in 3DDS Cd$_3$As$_2$ considering electron scattering by acoustic (ap) and optical (op) phonons. We follow the Boltzmann transport equation technique in the relaxation time approximation. The mobility is calculated as a function of temperature and electron density with and without the screening of electron-phonon coupling. Simple expressions are obtained for mobility in both the Bloch-Gruneisen and equipartition regimes. The calculated results are compared with the experiment.

## II. THEORY OF PHONON LIMITED MOBILITY
### A. Dirac-Fermions in three-dimensional Dirac semimetal

We consider 3D Dirac-Fermions in a 3DDS with a large Fermi energy $E_f$ in the conduction band so that the charge carriers are electrons. In a 3DDS, the electron energy dispersion is linear $E_\mathbf{k}=\hbar v_f k$ and the density of states is $D(E_\mathbf{k}) = gE_\mathbf{k}^2/[2\pi^2(\hbar v_f)^3]$, where $v_f$ is the Fermi velocity, $\mathbf{k}$ is the 3D wave vector, and $g = g_s g_v$ with $g_s$ ($g_v$) being the spin (valley) degeneracy of the electron. We assume, at this stage, that the electronic dispersion is isotropic [30,31], although it is found to be anisotropic by some authors [2,5,10]. In 3D Dirac semimetal Cd$_3$As$_2$, the electron density is generally high ($\sim 10^{17}$-$10^{19}$ cm$^{-3}$) and $E_f$ is sufficiently large ($\sim$ 75- 350 meV). Consequently, electrons with $E_\mathbf{k} \approx E_f$ will contribute to the transport properties, as in metals. Since, in Cd$_3$As$_2$ the optical phonon energy is about 25 meV [17], which is much smaller than the $E_f$, we assume the scattering by optical phonons to be quasi-elastic. Hence, we obtain phonon (both acoustic and optical) limited electron mobility using the semi-classical Boltzmann transport equation solved in the relaxation time approximation. Interestingly, it leads to simple expressions for the ap and op limited mobility.

The energy dependent momentum relaxation time for the scattering due to phonons is shown to be [30, 36-38]

$$\frac{1}{\tau(E_\mathbf{k})} = \left(\frac{V}{8\pi^3}\right)\left(\frac{1}{[1-f_o(E_\mathbf{k})]}\right)\int d^3k' [1-f_o(E_{\mathbf{k'}})]\left[1 - \frac{\mathbf{v}_{\mathbf{k'}} \cdot \mathbf{E}}{\mathbf{v}_\mathbf{k} \cdot \mathbf{E}}\right] W(\mathbf{k},\mathbf{k'}),$$

(1)

where $\mathbf{v}_\mathbf{k} = (1/\hbar)(dE_\mathbf{k}/d\mathbf{k}) = v_f \mathbf{k}/|\mathbf{k}|$ is the group velocity of the electron in the state $\mathbf{k}$, $\mathbf{E}$ is the electric field and $W(\mathbf{k}, \mathbf{k'})$ is the transition probability for electron scattering by phonons from state $\mathbf{k}$ to $\mathbf{k'}$. Substituting for $\mathbf{v}_\mathbf{k}$ and $\mathbf{v}_{\mathbf{k'}}$, [1- ($\mathbf{v}_{\mathbf{k'}} \cdot \mathbf{E}$)/($\mathbf{v}_\mathbf{k} \cdot \mathbf{E}$)] gives [1- (cos $\theta_{\mathbf{k'}}$ /cos $\theta_\mathbf{k}$)], where $\theta_\mathbf{k}$ and $\theta_{\mathbf{k'}}$ are, respectively, the angles of $\mathbf{k}$ and $\mathbf{k'}$ with the electric field. This is the same as for the elastic/quasi-elastic scattering. Interestingly, in 3DDS, [1- ($\mathbf{v}_{\mathbf{k'}} \cdot \mathbf{E}$)/($\mathbf{v}_\mathbf{k} \cdot \mathbf{E}$)] is independent of the magnitudes of $\mathbf{k'}$ and $\mathbf{k}$. We may express cos $\theta_{\mathbf{k'}}$ = cos $\theta_\mathbf{k}$ cos $\theta$ + sin $\theta_\mathbf{k}$ sin $\theta$ cos $\beta$, where $\theta$ is the angle between $\mathbf{k}$ and $\mathbf{k'}$, and $\beta$ is the angle between the planes containing **k** and **k′**, and **k** and **E**. In Eq.(1), none of the terms other than $\cos\theta_{\mathbf{k}'}$ contains $\beta$. Taking **k** as polar axis, integration with respect to $\beta$ from 0 to $2\pi$, appearing in $d^3k' = k'^2 dk' \sin\theta\, d\theta\, d\beta$, gives zero value for the second term in $\cos\theta_{\mathbf{k}'}$ [37]. Thus, we get

$$\frac{1}{\tau(E_\mathbf{k})} = \left(\frac{V}{8\pi^3}\right)\left(\frac{1}{[1-f_o(E_\mathbf{k})]}\right)\int d^3k'[1-f_o(E_{\mathbf{k}'})][1-\cos\theta]W(\mathbf{k},\mathbf{k}'). \quad (2)$$

The transition probability for the electron scattering by phonons of energy $\hbar\omega_\mathbf{q}$ and wave vector **q** is given by

$$W(\mathbf{k},\mathbf{k}') = \frac{2\pi}{\hbar}\sum_\mathbf{q}\frac{|C(q)|^2}{\varepsilon^2(q)}[N_\mathbf{q}\delta(E_\mathbf{k}-E_{\mathbf{k}'}+\hbar\omega_\mathbf{q})\delta_{\mathbf{k}',\mathbf{k}+\mathbf{q}} \quad (3)$$
$$+(N_\mathbf{q}+1)\delta(E_\mathbf{k}-E_{\mathbf{k}'}-\hbar\omega_\mathbf{q})\delta_{\mathbf{k}',\mathbf{k}-\mathbf{q}}\theta(E_\mathbf{k}-\hbar\omega_\mathbf{q})],$$

where $|C(q)|^2$ is the electron-phonon matrix element, $\varepsilon(q)$ is the screening function, $N_\mathbf{q} = [\exp(\hbar\omega_\mathbf{q}/k_BT)-1]^{-1}$ is the phonon distribution function and $\theta(x)$ is the step function. After substituting $W(\mathbf{k},\mathbf{k}')$ in Eq.(2), the momentum relaxation time is given by

$$\frac{1}{\tau(E_\mathbf{k})} = \left(\frac{V}{8\pi^3}\right)\left(\frac{2\pi}{\hbar}\right)\left(\frac{1}{[1-f_o(E_\mathbf{k})]}\right)\int d^3k'(1-\cos\theta)$$
$$\times [1-f_o(E_{\mathbf{k}'})]\sum_\mathbf{q}\frac{|C(q)|^2}{\varepsilon^2(q)}\times [N_\mathbf{q}\delta(E_{\mathbf{k}'}-E_\mathbf{k}-\hbar\omega_\mathbf{q})\delta_{\mathbf{k}',\mathbf{k}+\mathbf{q}} \quad (4)$$
$$+(N_\mathbf{q}+1)\delta(E_{\mathbf{k}'}-E_\mathbf{k}+\hbar\omega_\mathbf{q})\delta_{\mathbf{k}',\mathbf{k}-\mathbf{q}}\theta(x)].$$

The summation over **q** is carried out by applying the Kronecker delta-functions. Using the dispersion relation, integration over $k'$ is expressed in terms of $E_{\mathbf{k}'}$. Then, integration gives

$$\frac{1}{\tau(E_k)} = \left(\frac{V}{2\pi\hbar(\hbar v_f)^3}\right)[1-f_o(E_\mathbf{k})]^{-1}\int_0^\pi d\theta(1-\cos\theta)\sin\theta$$
$$\times \frac{|C(q)|^2}{\varepsilon^2(q)}\{N_\mathbf{q}(E_\mathbf{k}+\hbar\omega_\mathbf{q})^2[1-f_o(E_\mathbf{k}+\hbar\omega_\mathbf{q})] \quad (5)$$
$$+(N_\mathbf{q}+1)(E_\mathbf{k}-\hbar\omega_\mathbf{q})^2[1-f_o(E_\mathbf{k}-\hbar\omega_\mathbf{q})]\theta(x)\},$$

where $\varepsilon(q) = [1+(q_{TF}/q)^2]$ is the screening function and $q_{TF} = [4\pi e^2 D(E_f)/\varepsilon_s]^{1/2}$ is the Thomas–Fermi wave vector which is assumed to be independent of temperature for $T < T_f$ (the Fermi temperature) [30].

**1. Relaxation time due to acoustic phonons**

Electrons are assumed to interact with longitudinal acoustic phonons via deformation potential coupling. The corresponding matrix element is

$$|C(q)|^2 = \left(\frac{D^2\hbar\omega_q}{2\rho_m V v_s^2}\right)\left(\frac{1+\cos\theta}{2}\right), \quad (6)$$

where $D$ is the acoustic deformation potential coupling constant, $\rho_m$ is the mass density of the material and $v_s$ is the acoustic phonon velocity.

In the quasi-elastic approximation, $q = 2k\sin(\theta/2)$, which gives $\cos\theta = [1-2(q/2k)^2]$ and $\sin\theta\, d\theta = q\, dq/k^2$. Substituting for $|C(q)|^2$ in Eq. (5), we obtain

$$\left(\frac{1}{\tau_{ap}(E_k)}\right) = \left(\frac{D^2}{2\pi\hbar(\hbar v_f)\rho_m v_s^2}\right)[1-f_o(E_k)]^{-1}\int_0^{2k}dq\,q\left(\frac{q}{2k}\right)^2\left[1-\left(\frac{q}{2k}\right)^2\right]$$
$$\times \frac{1}{\varepsilon^2(q)}\hbar\omega_q\{N_\mathbf{q}[1-f_o(E_k+\hbar\omega_\mathbf{q})]+(N_\mathbf{q}+1)[1-f_o(E_k-\hbar\omega_\mathbf{q})]\}. \quad (7)$$

Then, for $\hbar\omega_\mathbf{q} \ll E_f$, using the following relations,

$$f_o(E_\mathbf{k})[1-f_o(E_\mathbf{k}+\hbar\omega_\mathbf{q})] \approx -\hbar\omega_\mathbf{q}(N_\mathbf{q}+1)[\partial f_o(E_\mathbf{k})/\partial E_\mathbf{k}], \quad (8a)$$
$$f_o(E_\mathbf{k})[1-f_o(E_\mathbf{k}-\hbar\omega_\mathbf{q})] \approx -\hbar\omega_\mathbf{q} N_\mathbf{q}[\partial f_o(E_\mathbf{k})/\partial E_\mathbf{k}], \quad (8b)$$
$$f_o(E_\mathbf{k})[1-f_o(E_\mathbf{k})] = -k_BT[\partial f_o(E_\mathbf{k})/\partial E_\mathbf{k}], \quad (8c)$$

we obtain

$$\left(\frac{1}{\tau_{ap}(E_\mathbf{k})}\right) = \left(\frac{D^2(\hbar v_f)}{4\pi\rho_m\hbar v_s^2 E_\mathbf{k}^2(k_BT)}\right)\int_0^{2k}dq\,q^3\left[1-\left(\frac{q}{2k}\right)^2\right] \quad (9)$$
$$\times \frac{1}{\varepsilon^2(q)}(\hbar\omega_q)^2 N_q(N_q+1).$$

This equation is valid for all the temperatures in the quasi-elastic approximation. Introducing a dimensionless variable $y = (\hbar\omega_\mathbf{q}/k_BT)$, with the upper limit of integration $y_m = (2\hbar v_s k/k_BT)$, we get

$$\left(\frac{1}{\tau_{ap}(E_k)}\right) = \left(\frac{D^2(\hbar v_f)(k_BT)^5}{4\pi\rho_m(\hbar v_s)^5 v_s E_\mathbf{k}^2}\right)\int_0^{ym}dy\,y^5\left[1-\left(\frac{y}{y_m}\right)^2\right]$$
$$\times \frac{1}{\varepsilon^2(y)}N(y)[N(y)+1], \quad (9a)$$

where $\varepsilon^2(y) = [1+\{q_{TF}/(\hbar v_s/k_BT)\}^2/y^2]^2$, and $N(y) = (e^y-1)^{-1}$. For the unscreened el-ap interaction $\varepsilon(y) = 1$. Since $n_e$ is very large in 3DDS $Cd_3As_2$, $k$ and $E_k$ will be taken at Fermi energy. In the very low temperature regime and in the equipartition regime, we get simple expressions for the acoustic phonon relaxation time.

In the Bloch-Gruneisen (BG) regime, i.e. for $T \ll T_{BG}$ ($=2\hbar v_s k_f/k_B$), the BG temperature, $q\to 0$, and $\hbar\omega_\mathbf{q} \approx k_BT$. Hence, in Eq. (9a), we set $k = k_f$, $E_k = E_f$ and $\varepsilon^2(q) \approx (q_{TF}/q)^4 = [\{q_{TF}/(\hbar v_s/k_BT)\}^2/y^2]^2$. As $T\to 0$, the upper limit $y_m = y_{mf} = (2\hbar v_s k_f/k_BT) \to \infty$. Then, in the BG regime the momentum relaxation time is given by

$$\frac{1}{\tau_{ap-BG}(E_k)} = \left(\frac{D^2(\hbar v_f)(k_BT)^9}{4\pi\rho_m v_s(\hbar v_s)^9 E_\mathbf{k}^2 q_{TF}^4}\right)\int_0^\infty dy\,y^9 N(y)(N(y)+1). \quad (10)$$

The integration gives an expression for the relaxation time due to the screened el-ap coupling

$$\left(\frac{1}{\tau_{ap-BG}}\right) = \left(\frac{D^2(\hbar v_f)(k_BT)^9 9!\zeta(9)}{4\pi\rho_m\hbar E_f^2(\hbar v_s)^9 v_s q_{TF}^4}\right). \quad (10a)$$

where $\zeta(n)$ is the Riemann zeta function. For the unscreened el-ap coupling, setting $\varepsilon(y) =1$, we obtain [39]

$$\left(\frac{1}{\tau_{ap-BG}}\right) = \left(\frac{D^2(\hbar v_f)(k_BT)^5 5!\zeta(5)}{4\pi\rho_m \hbar E_f^2 (\hbar v_s)^5 v_s}\right), \quad (10b)$$

We find the power laws $\tau_{ap-BG}(E_f) \sim E_f^6$ (i.e $n_e^2$) and $T^{-9}$ for the screened case, whereas $\tau_{ap-BG}(E_f) \sim E_f^2$ (i.e $n_e^{2/3}$) and $T^{-5}$ for the unscreened case.

In the equipartition (EP) regime (i.e. for $T > \sim 20$ K), $\hbar\omega_q << k_BT$ and $N_q+1 \approx N_q =(k_BT/\hbar\omega_q)$ [38,40,41]. Then, setting $N(y)[N(y)+1]= y^{-2}$, we get

$$\left(\frac{1}{\tau_{ap-EP}(E_\mathbf{k})}\right) = \left(\frac{D^2(\hbar v_f)(k_BT)^5}{4\pi\rho_m(\hbar v_s)^5 v_s E_\mathbf{k}^2}\right)\int_0^{y_m} dy y^3 \left[1-\left(\frac{y}{y_m}\right)^2\right]\frac{1}{\varepsilon^2(y)}. \quad (11)$$

For the unscreened el-ap interaction, the above equation gives a simple analytical result

$$\left(\frac{1}{\tau_{ap-EP}(E_\mathbf{k})}\right) = \left(\frac{D^2 k_B T E_\mathbf{k}^2}{3\pi\hbar(\hbar v_f)^3 \rho v_s^2}\right). \quad (12)$$

We notice that, in the EP regime, $\tau_{ap-EP}(E_\mathbf{k}) \sim E_\mathbf{k}^{-2}$ and $T^{-1}$ for the unscreened el-ap interaction.

**2. Relaxation time due to polar optical phonons**

The electrons in 3DDS are assumed to interact with the polar optical phonons via Frohlich coupling, with the matrix element

$$|C(q)|^2 = \left(\frac{C_o}{q^2}\right)\left(\frac{1+\cos\theta}{2}\right), \quad (13)$$

where $C_o = (2\pi e^2 \hbar\omega_o \varepsilon')/V$, $\hbar\omega_\mathbf{q}=\hbar\omega_o$ is the optical phonon energy, $\varepsilon' = (\varepsilon_\infty^{-1} - \varepsilon_s^{-1})$, and $\varepsilon_\infty$ ($\varepsilon_s$) is the high frequency (static) dielectric constant. Using the above matrix element in Eq.(5), in the process of integration, $q^2$ in the denominator and in the screening function is replaced by $q_+^2$ and $q_-^2$, respectively, in the phonon absorption and emission terms. The expressions for $q_+$ and $q_-$ are obtained using the momentum and energy conservations in the scattering process and are given by

$$q_+^2 = (1/\hbar v_f)^2[2E_\mathbf{k}^2 + 2E_\mathbf{k}\hbar\omega_o + (\hbar\omega_o)^2 - 2E_\mathbf{k}(E_\mathbf{k}+\hbar\omega_o)\cos\theta]$$
$$= (1/\hbar v_f)^2[2E_\mathbf{k}(E_\mathbf{k}+\hbar\omega_o)(1-\cos\theta)], \text{ for } E_\mathbf{k}>>\hbar\omega_o, \quad (14a)$$

and

$$q_-^2 = (1/\hbar v_f)^2[2E_\mathbf{k}^2 - 2E_\mathbf{k}\hbar\omega_o + (\hbar\omega_o)^2 - 2E_\mathbf{k}(E_\mathbf{k}-\hbar\omega_o)\cos\theta]$$
$$= (1/\hbar v_f)^2[2E_\mathbf{k}(E_\mathbf{k}-\hbar\omega_o)(1-\cos\theta)], \text{ for } E_\mathbf{k}>>\hbar\omega_o. \quad (14b)$$

After integrating with respect to $E_{k'}$, the relaxation time due to the polar optical phonons is given by

$$\left(\frac{1}{\tau_{op}(E_\mathbf{k})}\right) = \left(\frac{C_o V}{2\pi\hbar(\hbar v_f)}\right)\left(\frac{1}{[1-f_o(E_\mathbf{k})]}\right)\int_0^\pi d\theta[1-\cos\theta]\left[\frac{1+\cos\theta}{2}\right]\sin\theta$$
$$\times\left\{\begin{array}{l}N_\mathbf{q}\left(\frac{(E_\mathbf{k}+\hbar\omega_0)^2}{q_+^2\varepsilon^2(q_+)}\right)[1-f_0(E_\mathbf{k}+\hbar\omega_0)] \\ +(N_\mathbf{q}+1)\left(\frac{(E_\mathbf{k}-\hbar\omega_0)^2}{q_-^2\varepsilon^2(q_-)}\right)[1-f_0(E_\mathbf{k}-\hbar\omega_0)]\theta(x)\end{array}\right\}. \quad (15)$$

Substituting the values of $q_+$ and $q_-$, for $E_\mathbf{k} >>\hbar\omega_o$, we get

$$\left(\frac{1}{\tau_{op}(E_\mathbf{k})}\right) = \left(\frac{C_o V}{2\pi\hbar(\hbar v_f)}\right)\left(\frac{1}{f_o(E_\mathbf{k})[1-f_o(E_\mathbf{k})]}\right)\int_0^\pi d\theta\left[\frac{1+\cos\theta}{2}\right]\sin\theta$$
$$\times\left\{\begin{array}{l}N_\mathbf{q}\left(\frac{(E_\mathbf{k}+\hbar\omega_0)}{2E_\mathbf{k}\varepsilon^2(q_+)}\right)f_0(E_\mathbf{k})[1-f_0(E_\mathbf{k}+\hbar\omega_0)] \\ +(N_\mathbf{q}+1)\left(\frac{(E_\mathbf{k}-\hbar\omega)_0}{2E_\mathbf{k}\varepsilon^2(q_-)}\right)f_0(E_\mathbf{k})[1-f_0(E_\mathbf{k}-\hbar\omega_0)]\theta(x)\end{array}\right\}. \quad (16)$$

Making the same approximations as in Eqs. (8a-c), we get

$$\left(\frac{1}{\tau_{op}(E_\mathbf{k})}\right) = \left(\frac{C_o V}{2\pi\hbar(\hbar v_f)}\right)\left(\frac{\hbar\omega_o}{k_BT}\right)N_\mathbf{q}(N_\mathbf{q}+1)\int_0^\pi d\theta\left(\frac{1+\cos\theta}{2}\right)\sin\theta$$
$$\times\left\{\frac{(E_\mathbf{k}+\hbar\omega_o)}{2E_\mathbf{k}\varepsilon^2(q_+)}+\frac{(E_\mathbf{k}-\hbar\omega_o)}{2E_\mathbf{k}\varepsilon^2(q_-)}\theta(x)\right\}. \quad (17)$$

For the unscreened el-op interaction, $\varepsilon^2(q_\pm) =1$. For the electron density under consideration, $x = (E_f - \hbar\omega_o) >1$, and $\theta(x)=1$. Then,

$$\left(\frac{1}{\tau_{op}(E_\mathbf{k})}\right) = \left(\frac{C_o V}{2\pi\hbar(\hbar v_f)}\right)\left(\frac{\hbar\omega_o}{k_BT}\right)N_\mathbf{q}(N_\mathbf{q}+1). \quad (18)$$

Substituting for $C_o$, we obtain

$$\left(\frac{1}{\tau_{op}(E_\mathbf{k})}\right) = \left(\frac{e^2(\hbar\omega_o)\varepsilon'}{\hbar(\hbar v_f)}\right)\left(\frac{\hbar\omega_o}{k_BT}\right)N_\mathbf{q}(N_\mathbf{q}+1). \quad (18a)$$

Interestingly, for the unscreened el-op coupling, $\tau_{op}(E_\mathbf{k})$ is independent of the electron energy, i.e. $\tau_{op}(E_\mathbf{k}) = \tau_{op}$ and its temperature dependence comes through $(\hbar\omega_o/k_BT)\times N_\mathbf{q}(N_\mathbf{q}+1)$.

**3. Phonon limited mobility in 3DDS**

The phonon limited electrical conductivity $\sigma$ and mobility $\mu$ are obtained using [36]
$$\sigma = e^2 K_0 \text{ and } \mu = \sigma/n_e e, \quad (19)$$
where, for the 3DDS, $K_0$ is shown to be

$$K_0 = \frac{v_f^2}{3}\int dE_\mathbf{k} D(E_\mathbf{k})\tau_i(E_\mathbf{k})\left(-\frac{\partial f_o(E_\mathbf{k})}{\partial E_\mathbf{k}}\right), \quad (19a)$$

with $i= ap$ and $op$. Then, the acoustic phonon limited mobility $\mu_{ap}$ can be obtained, using Eq. (9) in Eq.(19), by numerical integration. The $\mu_{ap}$ thus obtained is applicable in the entire temperature range of interest with and without screening. In the evaluation of the mobility, $[-\partial f_0(E_k)/\partial E_k]$ is replaced by $\delta(E_k-E_f)$. In the BG and EP regimes, simple analytical expressions for $\mu_{ap}$ are obtained.

In the BG regime, mobility is given by

$$\mu_{ap-BG} = \frac{8e\rho_m(\hbar v_s)^9 v_s E_f^4 q_{TF}^4}{3n_e\pi\hbar^4 v_f^2 D^2 (k_BT)^9 9!\zeta(9)} \quad,\text{ with screening,} \quad (20a)$$

$$= \frac{8e\rho_m(\hbar v_s)^5 v_s E_f^4}{3n_e\pi\hbar^4 v_f^2 D^2 (k_BT)^5 5!\zeta(5)} \quad,\text{ without screening.} \quad (20b)$$

Thus, $T$ and $n_e$ dependence in BG regime are given by the power laws $\mu_{ap-BG} \sim T^{-9}$ ($T^{-5}$) and $n_e^{5/3}$ ($n_e^{1/3}$) for the screened (unscreened) el-ap coupling.

In the EP regime, the mobility due to the unscreened el-ap coupling is found to be

$$\mu_{ap-EP} = \frac{ev_f^2 g\hbar \rho v_s^2}{2\pi D^2 n_e k_BT}. \quad (21)$$

In the EP regime, the power laws are $\mu_{ap-EP} \sim T^{-1}$ and $n_e^{-1}$.

The mobility due to the el-op scattering, with screening, can be obtained by using Eq.(17) in Eq.(19). For the unscreened el-op scattering, a simple analytical result is given by

$$\mu_{op} = \frac{gE_f^2}{6\pi^2 n_e e\hbar(\hbar\omega_o)\varepsilon'} \times \left(\frac{k_BT}{\hbar\omega_o}\right) \times \frac{1}{N_q(N_q+1)}$$

$$= \left(\frac{g}{6\pi^2}\right)^{1/3} \times \frac{(\hbar v_f)^2}{n_e^{1/3} e\hbar(\hbar\omega_0)\varepsilon'} \times \left(\frac{k_BT}{\hbar\omega_0}\right) \times \frac{1}{N_q(N_q+1)} \quad (22)$$

For $(\hbar\omega_o/k_B) = \theta >> T$, $\mu_{op} \sim (T/\theta) e^{\theta/T}$, and for $\theta << T$, $\mu_{op} \sim \theta/T$. Moreover, in the EP regime, $\mu_{op} \sim n_e^{-1/3}$.

The resultant phonon limited mobility is obtained by using the Matthiessen's rule $\mu_{ph} = (\mu_{ap}^{-1} + \mu_{op}^{-1})^{-1}$.

**B. Highly degenerate three-dimensional electron gas in semiconductor**

For comparison, we give expressions for the mobility due to the acoustic and optical phonons for three-dimensional electron gas (3DEG) in a highly degenerate semiconductor (i.e. with large Fermi energy). The electron energy dispersion is assumed to be parabolic $E_k = (\hbar k)^2/2m$ (with $m$ being the effective mass of the electron) and the density of states is $D(E_k) = (g/\sqrt{2}\pi)(m/\hbar)^{3/2} E_k^{1/2}$. Again, we make use of the fact that, in these semiconductors, Fermi energy ($E_f$=100- 470 meV for $n_e = 10^{18} - 10^{19}$ cm$^{-3}$) is much larger than the acoustic and optical phonon energy, and the scattering is assumed to be quasi-elastic. Then, the mobility in 3DEG is also obtained in the momentum relaxation time approximation.

An expression for $\tau_{ap}$ in 3DEG is given by

$$\frac{1}{\tau_{ap}(E_k)} = \left(\frac{D^2(k_BT)^5}{2^{7/2}\pi\rho_m m^{1/2}(\hbar v_s)^4 v_s^2 E_k^{3/2}}\right) \int_0^{ym} dy \frac{1}{\varepsilon^2(y)} y^5 N(y)[N(y)+1], \quad (23)$$

where the screening function, for 3DEG, in the Thomas-Fermi approximation, is $\varepsilon(q) = [1+(q_{TF}/q)^2]$ with $q_{TF}$ of 3DEG [37,40].

An expression for $\tau_{op}$ in 3DEG is found to be

$$\frac{1}{\tau_{op}(E_k)} = \left(\frac{e^2(\hbar\omega_o)\varepsilon' m^{1/2}}{2^{3/2}\hbar^2}\right)\left(\frac{\hbar\omega_o}{k_BT}\right) N_q(N_q+1)$$

$$\times \int_0^\pi d\theta \sin\theta \left[\frac{1}{E_k^{1/2}\varepsilon^2(q_+)} + \frac{1}{E_k^{1/2}\varepsilon^2(q_-)}\theta(x)\right]. \quad (24)$$

The mobility, in highly degenerate 3DEG, is obtained, again, using Eq.(19). In order to evaluate mobility, $K_0$ for 3DEG is given by

$$K_0 = \frac{2}{3m}\int dE_k E_k D(E_k)\tau_i(E_k)\left(-\frac{\partial f_o(E_k)}{\partial E_k}\right). \quad (25)$$

Due to the large electron density, $[-\partial f_0(E_k)/\partial E_k]$ is replaced by $\delta(E_k-E_f)$, as in 3DDS.

In the BG regime, $\mu_{ap-BG}$ is found to be

$$\mu_{ap-BG} = \frac{2^5 e\rho_m m(\hbar v_s)^8 v_s^3 E_f^3 q_{TF}^4}{3\pi n_e \hbar^3 D^2 (k_BT)^9 9!\zeta(9)} \quad,\text{ with screening,} \quad (26a)$$

$$= \frac{2^5 e\rho_m m(\hbar v_s)^4 v_s^2 E_f^3}{3\pi n_e \hbar^3 D^2 (k_BT)^5 5!\zeta(5)} \quad,\text{ without screening.} \quad (26b)$$

Hence, we find the power laws $\mu_{ap-BG} \sim T^{-9}$ ($T^{-5}$) and $\mu_{ap-BG} \sim n_e^{5/3}(n_e)$ for the screened (unscreened) el-ap coupling in 3DEG.

A simple analytical result is obtained for $\mu_{ap-EP}$ in the EP regime for the unscreened el-ap interaction. It is given by

$$\mu_{ap-EP} = \frac{\pi^{1/3} e\hbar^3 \rho v_s^2}{3^{1/3} D^2 m^2 n_e^{1/3} k_BT}. \quad (27)$$

In 3DEG, we find $\mu_{ap-EP} \sim T^{-1}$ and $n_e^{-1/3}$.

The mobility due to the polar optical phonons, for the unscreened el-op interaction, is given by an analytical equation

$$\mu_{op} = \frac{3^{1/3}\pi^{2/3}\hbar^3 n_e^{1/3}}{2e(\hbar\omega_o)m^2\varepsilon'} \times \left(\frac{k_BT}{\hbar\omega_o}\right) \times \frac{1}{N_q(N_q+1)}, \quad (28)$$

which shows the same $T$ dependence as in 3DDS. Also, we find $\mu_{op} \sim n_e^{1/3}$.

**III. RESULTS AND DISCUSSION**

We present here the numerical calculations of the phonon limited mobility in 3DDS Cd$_3$As$_2$ as a function of temperature $T$ (1-300 K) and electron density $n_e$ ($\sim 10^{17}$ – $10^{19}$ cm$^{-3}$). The material parameters used are: $v_f$ = 1x10$^8$ cm/s, $\rho_m$= 7.0 gm/cm$^3$, $v_s$= 2.3x10$^5$ cm/s, $\varepsilon_\infty$ = 12, $\varepsilon_s$ =36, g = 4 and D = 20 eV, unless otherwise mentioned [42]. In the present calculation, we use a typical value for the optical phonon energy of 25 meV [17, 43]. Throughout the discussion $n_0$ =1x10$^{18}$ cm$^{-3}$ is used. For the mobility calculations of 3DEG in Cd$_3$As$_2$ semiconductor, m = 0.036 $m_0$ and g =2 [42] are used.

**A. Phonon limited mobility dependence on temperature and electron density**

The acoustic phonon limited mobility $\mu_{ap}$, optical phonon limited mobility $\mu_{op}$ and the resultant phonon limited mobility $\mu_{ph}$ are shown as a function of temperature, respectively, in Figs. 1(a-c) for $n_e$ = 0.5,1.0 and 3.0 $n_0$. These are shown with and without the screening.

In Fig.1a, we choose the temperature range 1-50 K in which $\mu_{ap}$ is dominant/ more significant. Screening is found to enhance $\mu_{ap}$. In the temperature region T > ~10 K), the effect of screening is about 20% for all $n_e$. As T decreases (for T < ~10 K), the effect of screening increases and it is different for different $n_e$. For example, at T= 1 K, screening enhances $\mu_{ap}$ by 33%, 40% and 64%, respectively, for $n_e$ = 0.5,1.0 and 3.0 $n_0$. The $\mu_{ap}$, with and without screening, decreases with increasing T. In the temperature range 5-300 K, the decrease of screened $\mu_{ap}$ for all the curves, as obtained by curve fitting, is given by $\mu_{ap}$ ~ $T^{-1}$. This is the same as in the unscreened case. It indicates that the screening does not affect the T dependence. Also, $\mu_{ap}$ ~ $T^{-1}$ behavior is not affected by $n_e$. In the low temperature region (T < ~ 5 K), $\mu_{ap}$ decreases rapidly with increasing T. Because of the additional temperature dependence coming from the screening function, in this temperature region (q ~ T), $\mu_{ap}$ with (without) screening decreases more (less) rapidly with increasing temperature. Finally, as T → 0, the decrease is expected to follow the BG power law $\mu_{ap-BG}$ ~ $T^{-9}$ ($T^{-5}$) for the screened (unscreened) el-ap interaction. The $\mu_{ap}$ is found to be smaller for larger $n_e$ for T > ~1-2 K.

In Fig 1b, $\mu_{op}$ is shown as a function of T in the range T = 20-300 K, where $\mu_{op}$ is significant/dominant. Screening is found to enhance $\mu_{op}$ by about 45%. It is the same over the entire temperature range of 20-300 K and for all $n_e$, indicating that the screening does not affect the T and $n_e$ dependence of $\mu_{op}$. The unscreened $\mu_{op}$ is found to decrease with increasing $T$ and it is expected to vary as $(T/\theta)[N_q(N_q+1)]^{-1}$. Therefore, the decrease of $\mu_{op}$ is more rapid at lower $T$ and less rapid at higher $T$. It is found that the variation of both screened and unscreened $\mu_{op}$ can be given by $\mu_{op} \sim T^{-\gamma}$, where $\gamma = 2.1$ for $T$=50-300 K. However, inclusion of still lower $T$ region will increase $\gamma$. For example, $\gamma = 2.37$ (3.7) for the range 40-300 (20-300) K. For $T << \theta$, the unscreened $\mu_{op} \sim (T/\theta) e^{\theta/T}$.

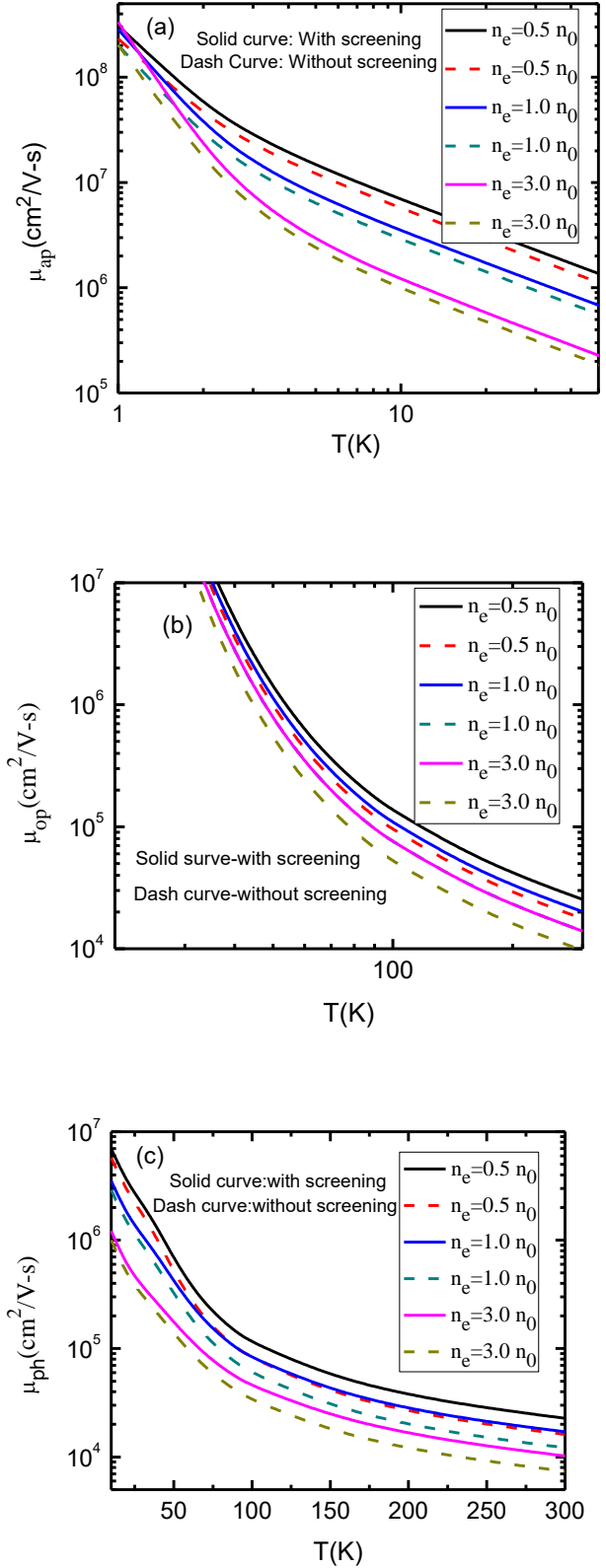

**FIG. 1.** Phonon limited mobility as a function of temperature for different electron density. Solid (dashed) curves are for screened (unscreened) electron-phonon interaction. (a) Mobility due to acoustic phonons $\mu_{ap}$, (b) mobility due to optical phonons $\mu_{op}$ and (c) resultant mobility due to acoustic and optical phonons $\mu_{ph}$.

The resultant phonon limited mobility $\mu_{ph}$ also decreases with increasing temperature (Fig. 1c). In the low $T$ region, the temperature dependence is mainly governed by the acoustic phonons. In the higher temperature region, it is largely governed by optical phonons. In the temperature range 10-300 K, the $\mu_{ph}$ variation can be given by $\mu_{ph} \sim T^{-\gamma}$ with $\gamma = 1.9, 1.75$ and $1.55$, respectively, for $n_e = 0.5, 1.0$ and $3.0$ $n_o$. This $n_e$ dependence of $\gamma$ may be attributed to different dominance and $n_e$ dependence of $\mu_{ap}$ and $\mu_{op}$, which we will see in the following. Screening is found to enhance the $\mu_{ph}$ and this enhancement is smaller at lower $T$ and larger at higher $T$.

In Figs. 2(a-c), the screened $\mu_{ap}$ and $\mu_{op}$, along with $\mu_{ph}$ are shown as a function of $T$ (10- 300 K) for $n_e$= 0.5, 1 and 3 $n_0$, respectively. Because of its stronger temperature dependence $\mu_{op}$ decreases more rapidly than $\mu_{ap}$ does in the temperature range considered. Moreover, $\mu_{op}$ dominates $\mu_{ap}$ at higher $T$ and a crossover is observed. The crossover temperature $T_c$ shifts to higher values as $n_e$ increases. For instance, for screened (unscreened) el-ph coupling, $T_c$= 50 (48) K, 57 (55) K and 77(70) K, respectively, for $n_e = 0.5, 1$ and 3 $n_0$. For a given $n_e$, for screened interaction, $T_c$ is slightly larger than that for the unscreened case, which may be attributed to the larger enhancement of $\mu_{op}$ as compared to the enhancement of $\mu_{ap}$. Moreover, $T_c$ also shifts to higher value for larger $D$, if it is varied, because $\mu_{ap} \sim D^{-2}$. Similar shift in the crossover temperature is found in electron cooling in 3DDS [35]. The total phonon limited mobility $\mu_{ph}$ is decreasing with increasing $T$, in qualitative agreement with the observed behavior and magnitude [7,12,23]. We have carried out calculations for some experimental samples and a quantitative comparison with the experimental results is made later in this section.

The temperature dependence of the phonon limited mobility in 3DDS and in highly degenerate 3DEG is found to be the same, in both BG and EP regimes. This may be attributed to the 3D nature of the phonons. But, in 3DEG the crossover temperature $T_c$, for the same $n_e$, is higher than that in 3DDS. This is because of different $n_e$ dependence in 3DDS and 3DEG. Moreover, for a given $n_e$, $\mu_{ph}$ in 3DDS is greater than that in 3DEG by about four times.

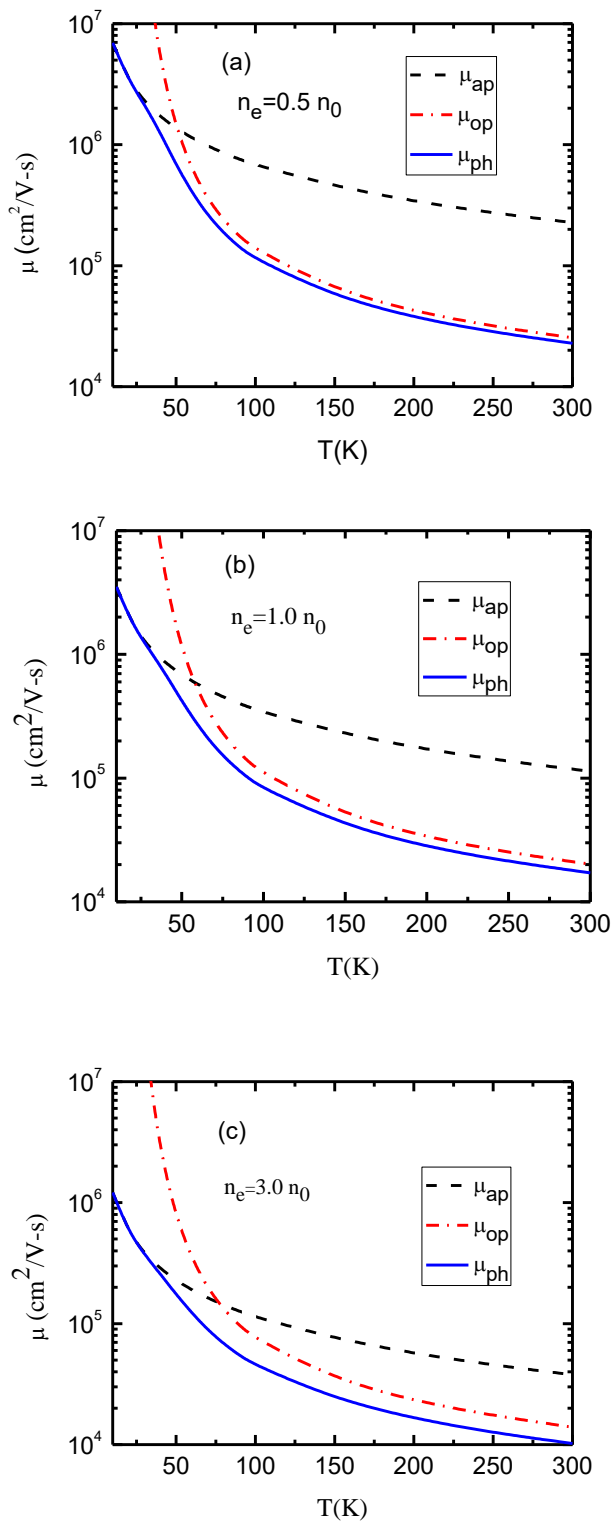

**FIG. 2.** Phonon limited mobility as a function of temperature due to the screened electron-phonon interactions for different electron density. (a) $n_e$= 0.5 $n_0$, (b) $n_e$= 1.0 $n_0$ and (c) $n_e$= 3.0 $n_0$.

The electron density dependence of the screened $\mu_{ap}$, $\mu_{op}$ and $\mu_{ph}$ for $n_e = 0.5-10\ n_0$ is shown in Figs. 3(a-c) for $T= 50, 100$ and $150$ K, respectively. We see that, for the temperatures chosen here (EP regime), $\mu_{ap}$ decreases as $n_e$ increases, more rapidly than $\mu_{op}$ does. At all the three temperatures, we find $\mu_{ap} \sim n_e^{-1}$ and $\mu_{op} \sim n_e^{-1/3}$. There is a cross over of $\mu_{ap}$ and $\mu_{op}$. The cross over electron density $n_{ec}$ shifts to higher value with increasing temperature. For example, for $T= 50, 100$ and $150$ K, $n_{ec}= 0.5\ (0.7)$, $6.0\ (7.5)$ and $9.5\ (13)\ n_0$, respectively, for the screened (unscreened) el-ap coupling. It is to be noted that the crossover value of $n_{ec}$ and the dominancy are expected to change when $D$ is varied. Moreover, we observe that $\mu_{ph}$ also decreases with increasing $n_e$, more rapidly at lower $T$. This may be attributed to the dominance of $\mu_{ap}$, which has stronger $n_e^{-1}$ dependence as compared to the $n_e^{-1/3}$ dependence of $\mu_{op}$. The $\mu_{ap} \sim n_e^{-1}$, found above, agrees exactly with the prediction in the EP regime for the unscreened case. The $\mu_{ph}$ curve gives $n_e^{-\beta}$ dependence with $\beta = 0.82$, $0.57$ and $0.52$, respectively, for $T=50, 100$ and $150$ K. This $T$ dependence of $\beta$ is due to the decreasing significance of $\mu_{ap}$ as $T$ increases. Interestingly, in the BG regime, $\mu_{ap-BG}$ increases with increasing $n_e$ and it follows the power law $\mu_{ap-BG} \sim n_e^{5/3}\ (n_e^{1/3})$ with (without) screening.

In 3DEG, in the BG regime, $\mu_{ap-BG} \sim n_e^{5/3}\ (n_e)$ with (without) screening as compared to the $\mu_{ap-BG} \sim n_e^{5/3}\ (n_e^{1/3})$ in 3DDS. The power law, with screening, $\mu_{ap-BG} \sim n_e^{5/3}$ in both 3DDS and 3DEG is a surprising result, although the density of states are different. In 3DEG, in the EP regime, for screened and unscreened el-ap coupling, the power laws are $\mu_{ap-EP} \sim n_e^{-1/3}$ and $\mu_{op} \sim n_e^{1/3}$, as compared to the $\mu_{ap-EP} \sim n_e^{-1}$ and $\mu_{op} \sim n_e^{-1/3}$ in 3DDS. Consequently, in 3DEG we find that, at a given $T$, in the EP regime the $\mu_{ph}$ weakly depends on $n_e$. The difference in the $n_e$ dependence of mobility in 3DDS and 3DEG may be attributed to the different $n_e$ dependence of the density of states. This difference in the $n_e$ dependence of the phonon limited mobility may be exploited to identify the Dirac phase of $Cd_3As_2$ from the experimental measurements. Secondly, it can be used to know the significance of the screening. The $T$ and $n_e$ dependence of the screened and unscreened $\mu_{ap}$ and $\mu_{op}$ in 3DDS and 3DEG, for both the BG and EP regimes, are given in Table I.

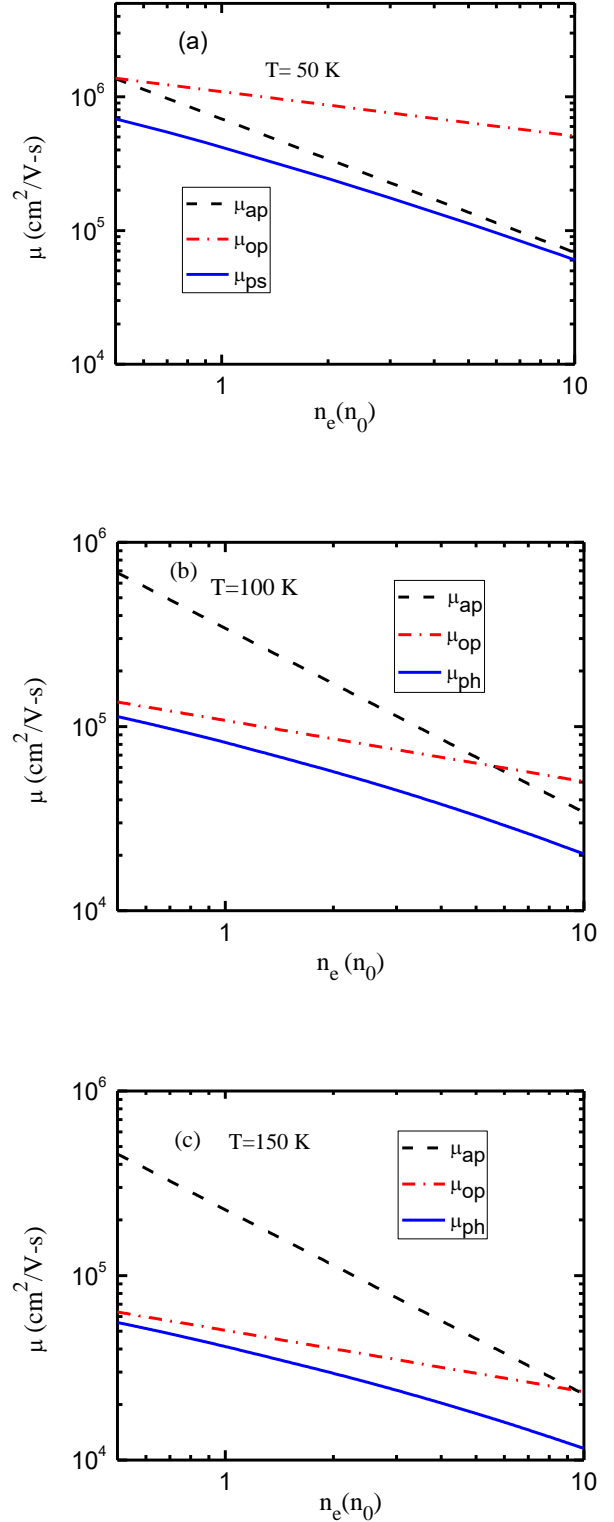

FIG. 3. Phonon limited mobility as a function of electron density due to the screened electron-phonon interaction at different temperature. (a) $T=50$ K, (b) $T=100$ K and (c) $T=150$ K.

**B. Comparison with experiment**

In the following, the mobility is calculated for the samples of Refs.[7, 12, 23] and detailed comparison is made with the experimental data. In order to obtain a better agreement with the experimental results we vary $D$, in the acoustic phonon limited mobility, which is in the range 10-30 eV [31, 42]. In the samples of Refs. [7, 12, 23], the very low temperature mobility /resistivity is governed by the impurity scattering. It is found to be nearly independent of the temperature and is called the residual mobility $\mu_I$ /resistivity $\rho_I$. We have used the observed residual mobility together with the calculated $\mu_{ph}$ to obtain agreement between the theoretical resultant mobility $\mu_T = (\mu_I^{-1} + \mu_{ph}^{-1})^{-1}$ and the experimental data.

In Fig. 4a, we have presented the calculated $\mu_T$, with the screened ap and op coupling, as a function of temperature for the sample of Cao et al. [12], and compared with their experimental data. In the sample of Cao et al., the residual mobility is $\mu_I = 1.87 \times 10^5$ cm$^2$/ V- s at 4.0 K. In the low temperature region, the observed mobility is nearly constant and then decreases with increasing $T$. Also, $n_e$ is very slowly decreasing (1.73 -1.57$\times 10^{18}$ cm$^{-3}$) with increasing $T$. We have calculated $\mu_{ph}$, using these electron densities at different $T$, and obtained the resultant mobility $\mu_T$. The curves are shown for $D$= 5, 10, 15 and 20 eV. The $\mu_T$ with $D$= 20 eV is much smaller than the observed mobility in the entire range of $T$. The $\mu_T$ curves with $D$= 10 and 15 eV are in reasonably good agreement with the experimental data for $T$< ~ 40 K, where $\mu_{ap}$ is dominant in $\mu_{ph}$. In the higher $T$ ( > ~ 50 K ) region the calculated values are about 1.2- 1.5 times smaller than the experimental values. Further reducing the value of $D$ (say up to 5 eV) does not help to obtain better agreement at higher $T$ region as $\mu_{ap}$ is less significant compared to $\mu_{op}$ in this region. Secondly, the choice of $D$= 5 eV is found to over estimate the calculated mobility in low temperature region where ap scattering is dominant. In Fig 4b, the calculated mobility $\mu_T$ with ap (op) coupling screened (unscreened), is shown along with the experimental data. There is a good agreement between the theoretical and experimental values for $T$ < ~ 30 K with $D$= 15 eV. However, for $T$ > ~ 50 K, the experimental values are about 1.5-2.0 times greater than the calculated values. We observe that the screened op scattering improves the agreement with the experimental values in the higher temperature region to some extent, as the $\mu_{op}$ enhances significantly due to the screening.

The theoretical calculations of $\mu_T$, with the screening of ap and op scattering, as a function of temperature are presented for the sample of Pariari et al

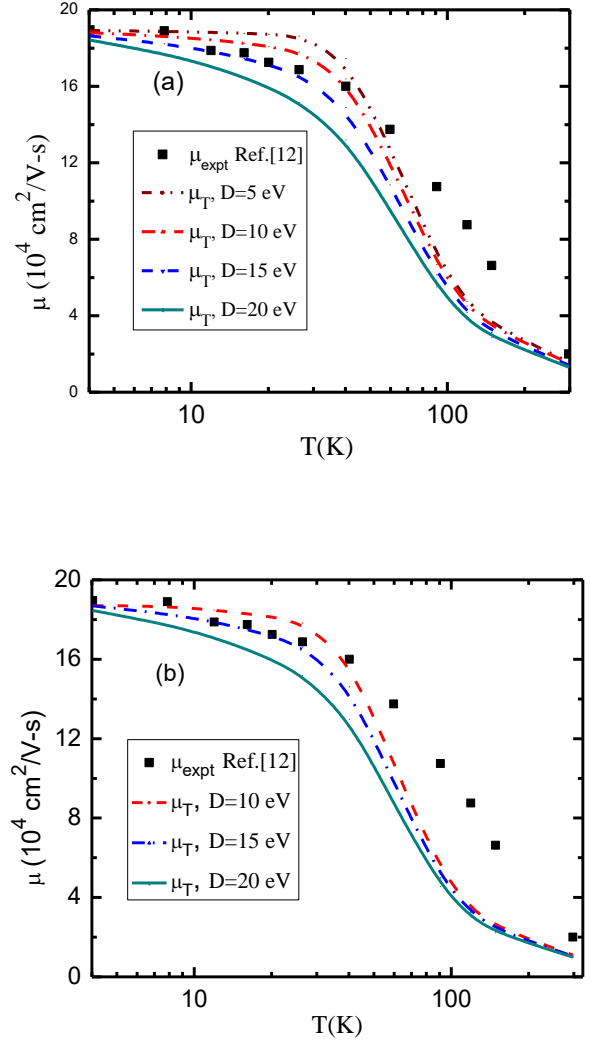

**FIG. 4.** Mobility as a function of temperature for the sample of Ref. [12]. (a) Screened electron acoustic and optical phonon interactions and (b) screened electron-acoustic phonon and unscreened electron-optical phonon interactions

[23] and compared with their experimental data in Fig 5a. Their measurements are $\rho$ as function of $T$ (2-350 K) for the sample with $n_e$ = 6.8 $n_0$, and having the residual resistivity $\rho_I$ =70.612 $\mu\Omega$-cm and residual mobility $\mu_I$ = 1.3 $\times$ 10$^4$ cm$^2$/ V- s at 2 K. We have converted their experimental $\rho$ values into mobility using $\mu = 1/(n_e e\rho)$. Although the experimental data is shown by a continuous curve in Ref.[23], we have chosen sufficiently good number of experimental values for comparison. Calculated values of $\mu_T$ are shown for $D$=20, 25, 30 and 35 eV. The $\mu_T$ values with $D$ = 30 eV are in good agreement with the experimental values in the

temperature region $T < \sim 100$ K and are larger in the region $T > \sim 100$ K. On the other hand, the $\mu_T$ values with $D = 35$ eV are reasonably agreeing with the observed values for $\sim 70$ K $< T < \sim 200$ K. $\mu_T$ with $D = 35$ eV underestimates the theoretical values for $T < \sim 50$ K and overestimates for $T > 200$ K.

In Fig. 5b, the $\mu_T$ curves with the unscreened op scattering are shown for $D = 20, 25$ and $30$ eV. The $\mu_T$

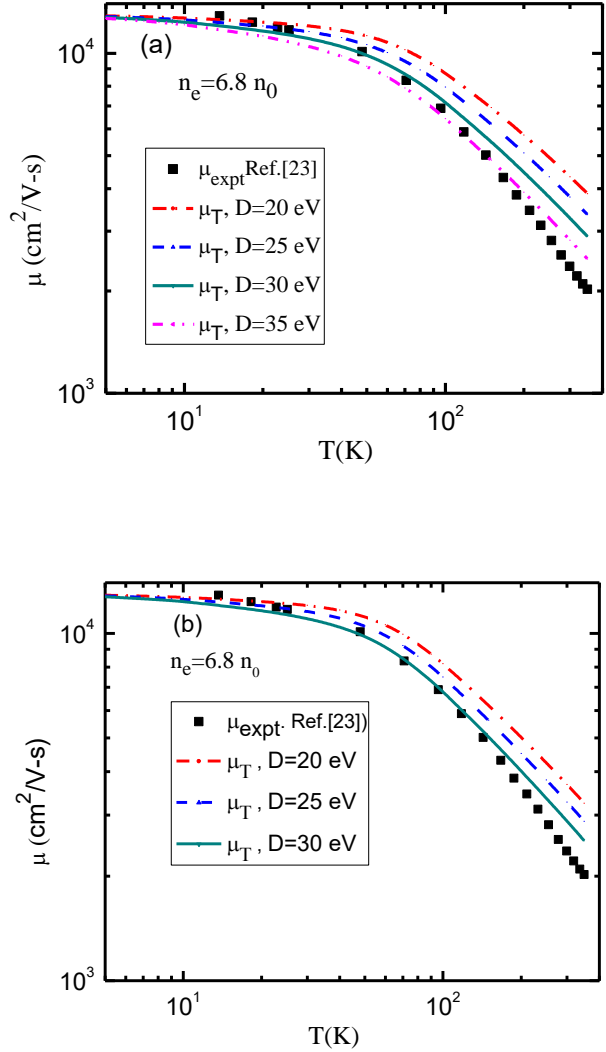

**FIG. 5.** Mobility as a function of temperature for the sample of Ref. [23]. (a) Screened electron acoustic and optical phonon interactions and (b) screened electron-acoustic phonon and unscreened electron-optical phonon interactions

curve with $D = 30$ eV is agreeing with the experimental values for $T < \sim 200$ K, and is marginally larger for $T > \sim 200$ K.

He et al. [7] have measured resistivity as a function of temperature, at zero magnetic field, in a sample with $n_e \approx 5.3\ n_0$. They observe a very flat curve for $T < 10$ K with a residual resistivity 28.2 $\mu\Omega$-cm and metallic behavior for $T > 10$ K. We have converted their resistivity data to mobility. By taking residual mobility $\mu_I = 4.18 \times 10^4$ cm$^2$/V-s, we have calculated mobility $\mu_T$ as a function of temperature and compared it with the experimental data in Fig. 6. Our calculations are presented with the screened ap and op coupling for $D = 10, 15$ and $20$ eV. We find a good agreement between theoretical and experimental values for $D = 15$ eV. We also note that, the curve due to $D = 20$ eV is equally closer to the experimental points but for the region of $T < \sim 30$ K.

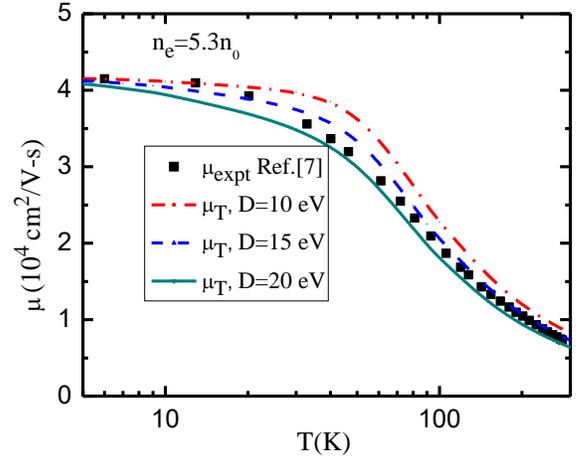

FIG. 6. Mobility as a function of temperature for the sample of Ref. [7] for the screened electron acoustic and optical phonon interactions.

In the experimental data of Ref. [13], the residual resistivity is observed for $T < 6$ K, and at this temperature the observed $\rho_I$ (11.6 n$\Omega$-cm) is giving a very large value of $\mu_I \sim 1 \times 10^8$ cm$^2$/V-s. At $T \sim 10$ K, we find a large $\mu = 1.78 \times 10^7$ cm$^2$/V-s from their measured $\rho$. Our calculated $\mu_{ap}$ (with screening) at $T = 10$ K, for $n_e = 5.86\ n_0$ (estimated from the SdH oscillations) also gives a large value $1.0$ $(0.26) \times 10^7$ cm$^2$/V-s for $D = 5$ (10) eV. Screened $\mu_{ap}$ for $D = 5$ eV is little closer to the observed value but still about 1.8 times smaller. However, for $T > \sim 10$ K, the experimental data shows $1/T$ behavior, which is satisfied by only $\mu_{ap}$, where as calculated $\mu_{ph} \sim T^{-\gamma}$ with $\gamma \sim 1.5$-$2.0$. We would like to point out that, in a recent work [Ref.44], the authors have used hypothetical temperature dependent mobility relations $\mu_{ap}^{-1} = A1 \times T^{3/2}$ and $\mu_{op}^{-1} = A2 \times N_q$, corresponding to a non-degenerate 3DEG [41], to explain their mobility data of 3DDS $Cd_3As_2$, in the range 20-350 K, for the sample with $n_e \sim 10 \, n_0$. These relations are contrary to our predictions $\mu_{ap}^{-1} \sim T$ and $\mu_{op}^{-1} \sim (\theta/T)[N_q(N_q+1)]$ for the large electron density 3DDS $Cd_3As_2$. The behavior of the resultant phonon limited mobility, in our model, gives $\mu_{ph}^{-1} \sim T^\gamma$ with $\gamma$=1.5-2.0.

In view of the above discussed comparisons between the theoretical and experimental results, we believe that more exclusive measurements of the intrinsic mobility with $T$ and $n_e$ dependence may throw more light on the understanding of el-ph interaction.

In the following, we would like to make a few remarks. Our simple theoretical model of el-ap coupling via deformation potential and el-op coupling due to Frohlich interaction with one optical phonon branch, although there can be many optical branches, is providing analytical results for the mobilities, and giving good agreement with the experimental results. In the absence of any other experimental work on el-op coupling strength in 3DDS $Cd_3As_2$, we have used the simple model of earlier work in $Cd_3As_2$ semiconductor [42] with modification. Possible reason for achieving good agreement with only one optical branch is that the LO phonon coupling via Frohlich interaction, among all other possible optical phonon contributions, may be stronger and dominant. It may be noted that in SnSe, scattering by the highest LO phonons, amongst many optical branches, is shown to be dominating the mobility [45]. The success of our present analytical theory lies in explaining the observed mobility with only deformation potential constant as a fitting parameter.

We point out that, to improve future calculations, the first principles calculations using a density functional theory (DFT) formalism (e.g., see Refs. [46,47]) can be applied to obtain the electronic band structure, phonon dispersion and el-ph coupling matrix elements for all phonon branches. The el-ph couplings, thus obtained, can be used as inputs in the Boltzmann transport equation to calculate the phonon limited mobility. These first principles calculations using DFT formalism may also address the anisotropy of electron band dispersion. Our mobility calculations with the simple isotropic band structure, still giving good agreement with the experimental results, may be closer to the geometric mean mobility [10]. It may also indicate that the band structure is nearly isotropic [3].

## IV. SUMMARY

The electron mobility due to acoustic $\mu_{ap}$ and optical $\mu_{op}$ phonon scattering is studied theoretically as a function of temperature $T$ and electron density $n_e$ in a relatively high electron density 3DDS $Cd_3As_2$. Screening of electron-phonon interaction is found to enhance mobility significantly. In the Bloch-Gruneisen regime, the temperature dependence of acoustic-phonon limited mobility is $\mu_{ap-BG} \sim T^{-9}$ ($T^{-5}$) and the electron density dependence is $\mu_{ap-BG} \sim n_e^{5/3}$ ($n_e^{1/3}$) for the screened (unscreened) el-ph interaction. In the equipartition regime, $\mu_{ap-EP} \sim T^{-1}$ and $n_e^{-1}$, and $\mu_{op} \sim (T/\theta)/[N_q(N_q+1)]$ and $n_e^{-1/3}$, where $N_q$ is the optical phonon distribution function. These differing $T$ and $n_e$ dependencies have led to the crossover of $\mu_{ap}$ and $\mu_{op}$. At a given $T$, mobility can be enhanced by tuning $n_e$. Our theoretical values are compared with the recent experimental results. By varying only the acoustic deformation potential constant, in the range 10-30 eV, a reasonably good agreement is obtained. In conventional highly degenerate 3DEG, the $T$ dependence is found to be the same as in 3DDS. However, the $n_e$ dependence differs with $\mu_{ap-BG} \sim n_e^{5/3}$ ($n_e$) with (without) screening and $\mu_{ap-EP} \sim n_e^{-1/3}$ and $\mu_{op} \sim n_e^{1/3}$. We suggest that the $n_e$ dependent measurements of the mobility in disorder free samples may be exploited to identify the Dirac phase of the electrons in $Cd_3As_2$.

**Table I:** Table showing the temperature $T$ and electron density $n_e$ dependence of the $\mu_{ap}$ and $\mu_{op}$ in BG and EP regime. In this table $f_{op}(T)= (T/\theta)[N_q(N_q+1)]^{-1}$, where $\theta = \hbar\omega_o/k_B$.

| $\mu$ | BG regime | | | | EP regime | | | |
|---|---|---|---|---|---|---|---|---|
| | 3DDS | | 3DEG | | 3DDS | | 3DEG | |
| $\mu_{apscr}$ | $T^{-9}$ | $n_e^{5/3}$ | $T^{-9}$ | $n_e^{5/3}$ | $T^{-1}$ | $n_e^{-1}$ | $T^{-1}$ | $n_e^{-1/3}$ |
| $\mu_{apunscr}$ | $T^{-5}$ | $n_e^{1/3}$ | $T^{-5}$ | $n_e$ | $T^{-1}$ | $n_e^{-1}$ | $T^{-1}$ | $n_e^{-1/3}$ |
| $\mu_{op}$ | N A | | | | $f_{op}(T)$ | $n_e^{-1/3}$ | $f_{op}(T)$ | $n_e^{1/3}$ |